# Molecular Beam Epitaxy Growth and Doping Modulation of Topological Semimetal NiTe$_2$

Liguo Zhang[1, a)], Dapeng Zhao[1], Xiangyang Liu[2], Junwen Lai[2], Junhai Ren[1], Qin Wang[1], Haicheng Lin[1, a)], Yan Sun[2, a)], Katsumi Tanigaki[1]

[1]Beijing Academy of Quantum Information Sciences, Beijing 100193, China

[2]Institute of Metal Research, Chinese Academy of Sciences, Shenyang 110016, China

[a]Authors to whom correspondence should be addressed: zhanglg@baqis.ac.cn, linhc@baqis.ac.cn and sunyan@imr.ac.cn

In this study, high-quality thin films of the topological semimetal phase NiTe$_2$ were prepared using molecular beam epitaxy (MBE) technique, confirmed through X-ray diffraction with pronounced Laue oscillations. Electrical transport experiments reveal that thick films have properties similar to bulk materials. By employing co-deposition, we introduced either magnetic or non-magnetic elements during the growth of thinner films, significantly altering their electrical properties. Notably, magnetic element Cr induces long-range ferromagnetic ordering, leading to the observation of significant anomalous Hall effect in NiTe$_2$ thin films. The Hall conductivity remains nearly constant well below the Curie temperature, indicating the correlation with the intrinsic topological nature of the band structure. Theoretical first principles band calculations support the generation of the Weyl semimetal state in the material through magnetic doping. These findings pave the way for exploring more magnetic Weyl semimetal materials and related low-dimensional quantum devices based on topological semimetal materials.

Transition metal dichalcogenides (TMDs) exhibit diverse crystal and electronic structures, leading to excellent physical properties, making them highly sought-after for future development in electronic devices[1-3]. Among them, 1$T$-NiTe$_2$ is a prototypical layered material categorized to type-II topological semimetals, hosting the Dirac points near the Fermi surface and chiral anomaly, which is ideal for studying topological physics[4-7]. Theoretical calculations suggest that $1T$-NiTe$_2$ exhibits attractive magnetic properties in its monolayer[8] and that its band structure can be tuned via substrate strain[9], pressure[10], or doping, potentially also leading to superconductivity[11]. The presence of topological states is crucial for the nontrivial topological superconductivity, sparking significant experimental interest[11,12]. Additionally, NiTe$_2$ promises in van der Waals (vdW) heterostructures, such as serving as a barrier layer for Cooper pairs in Josephson junctions[13,14]. Recent advances include the synthesis and study of bulk NiTe$_2$ and its doped counterparts[4,7,15,16], revealing superconducting signatures[15], topological band structures[4,7], etc. Moreover, strides have been taken towards achieving its low-dimensional forms, including nanowires and flakes[17-19]. The flakes prepared through chemical vapor deposition have showed solid stability in ambient conditions[18], making it an attractive candidate for constructing vdW heterostructure.

Considering the promising potential of low-dimensional NiTe$_2$, it is crucial to explore high-quality and controllable materials. Molecular beam epitaxy (MBE) provides precise control over thickness, composition, and carrier concentration. However, ideal preparation conditions for NiTe$_2$ are not yet fully researched[6,20]. Additionally, obtaining ultra-thin layers through mechanical exfoliation of single crystals remains challenging. These issues hinder a comprehensive understanding of the properties of ultrathin NiTe$_2$. The recent progress in synthesizing pure $1T$-NiTe$_2$ thin films on SiC substrates by MBE reveals essential differences between thin layers and bulk materials[6]. Moreover, MBE is well-suited for studying the doping effects. Especially through the method of magnetic doping, the introduction of ferromagnetic order into topological insulators breaks time-reversal symmetry and results in phenomena such as the quantum anomalous Hall effect and Axion insulator states[21,22]. Naturally, introducing ferromagnetic order by introducing magnetic elements could transform topological semimetals into magnetic Weyl semimetals, and thereby offer large possibilities to expand the current research on magnetic Weyl semimetal materials[23,24]. In this work, we utilized MBE to fabricate 1$T$-NiTe$_2$ thin films, investigating transport properties and achieving controlled elemental doping modulation. The observed long-range ferromagnetic order and anomalous Hall effect originating from the electronic band structure are further validated through theoretical calculations. Our findings may contribute to the future device applications based on ultra-thin NiTe$_2$.

NiTe$_2$ can exist in various distinct phases in low dimensions[6]. Among them, 1$T$-NiTe$_2$ is a layered material with a crystal structure belonging to the $P$-3$m$1 space group. Its monolayer unit has a sandwich-like structure, with a nickel atom layer sandwiched between two tellurium layers. The lattice parameters for 1$T$-NiTe$_2$ phase are $a = b =$ 3.843 Å and $c =$ 5.265 Å, as shown in Figs.1(a) and (b), depicting the top and side views. We employed an MBE co-deposition method to prepare the 1$T$-NiTe$_2$ thin films in an ultra-high vacuum chamber (pressure below 2×10$^{-9}$ Torr) on a high-temperature treated SrTiO$_3$(111) substrate. The reflection high-energy electron diffraction (RHEED) pattern in Fig.1(c) attests to the high quality of the substrate surface with structure reconstruction. High-purity (99.999%) Ni and Te sources were used, with a Ni:Te flux ratio of approximately 1:10 to ensure Te-rich condition and a growth rate of about 0.6 nm/min. The substrate temperature was maintained at 360°C during the growth, followed by a 30-minute annealing process. The RHEED pattern in Fig. 1(d) shows distinct stripes, indicating an in-plane crystal constant of about 3.85 Å (see supplementary material Fig. S1), and confirms the high film quality. X-ray Diffraction (XRD) analysis, shown in Fig. 1(e), displays a single preferred orientation with no other phases. The diffraction peaks indicate a crystal constant of about 5.284 Å along $c$-axis and an orientation of (0 0 0 1). The XRD curve also shows Laue oscillations around the (0 0 0 2) peak, suggesting high film quality. A zoomed 2$\theta$ view around this peak, scanned with longer acquisition time, is shown in Fig. 1(f). The data is simulated (red solid line) according to the equation[25]:

$$I(Q) = \left(\frac{\sin\left(\frac{QNc}{2}\right)}{\sin\left(\frac{Qc}{2}\right)}\right)^2 \quad (1)$$

where Q = 4πsin(θ)/λ is the reciprocal space vector, $N$ is the number of unit cells along the out-of-plane direction and $c$ is the lattice constant. The fitting curve is simulated considering peak position 2θ = 33.914° and thickness of 195.5 Å ($N$ = 37).

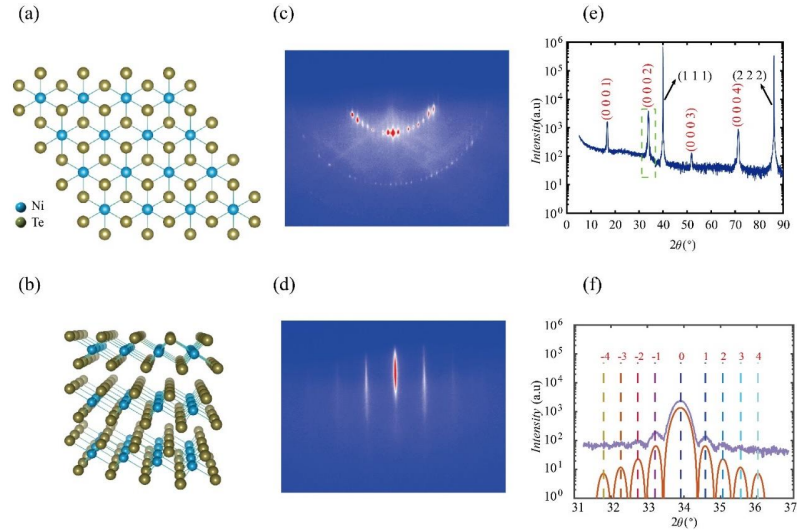

Fig. 1 (a)Top and (b) side view of the crystal structure of 1$T$-NiTe$_2$. (c-d) RHEED patterns of SrTiO$_3$ (111) surface (c) and 1$T$-NiTe$_2$ thin film (d) taken along the [11$\bar{2}$] direction of the substrate. (e) 2θ-ω XRD pattern of 1$T$-NiTe$_2$ thin film. (f) XRD pattern around the peak (0 0 0 2), and the red solid line represents the stimulated Laue oscillations, with corresponding dash lines indicating the indices of the Laue oscillation.

We first measured the temperature dependent resistivity ($RT$) of a 19.55 nm thick NiTe$_2$ film. The $RT$ curve showed a metallic behavior and is fitted well with the Bloch-Grüneisen (BG) formula[16]:

$$\rho(T) = \rho_0 + \alpha \left(\frac{T}{\theta_D}\right)^n \int_0^{\frac{\theta_D}{T}} \frac{x^n}{(e^x-1)(1-e^{-x})} dx \quad (2)$$

where the first term, $\rho_0$ represents temperature-independent residual resistivity while the second term accounts for the electric interactions. In the second term, $\alpha$ is a constant related to interaction strength and $\theta_D$ is the Debye temperature. We found that $n=2.8$ close to $n=3$, providing a well-fitted result akin to its bulk behavior[16]. It indicates that the resistivity of NiTe$_2$ thin film is dominated by the $s$-$d$ electron–phonon scattering[16]. The corresponding $\theta_D$ is approximately 187 K, lower than that of its bulk material reported as 230 K, being indicative of the reduced dimensionality[26]. Remarkably, the film resistivity is below 30 $\mu\Omega\cdot$cm across the entire temperature range, comparable to the previously reported value for a 50 nm thick film grown on a GaAs substrate via MBE[20], indicating high quality of the film. Dot plots in Figs. 2(b) and (c) illustrate sheet resistivity ($R_\square$) and Hall resistivity versus magnetic field ($B$) curves at 2 K for the 19.55 nm film. Hall measurements exhibited multiband conductivity, resembling those of NiTe$_2$ bulk materials and some other topological semimetals[27,28]. Assuming the contribution from two-carrier model, expressed as[28]:

$$\rho_{yx} = \frac{1}{e} \frac{(n_h \mu_h^2 - n_e \mu_e^2) + \mu_h^2 \mu_e^2 B^2 (n_h - n_e)}{(n_h \mu_h + n_e \mu_e)^2 + \mu_h^2 \mu_e^2 B^2 (n_h - n_e)^2} B \quad (3)$$

$$\rho_{xx} = \frac{(n_h \mu_h + n_e \mu_e) + (n_e \mu_e \mu_h^2 + n_h \mu_h \mu_e^2) B^2}{(n_h \mu_h + n_e \mu_e)^2 + \mu_h^2 \mu_e^2 B^2 (n_h - n_e)^2} \quad (4)$$

where $n_e$ ($n_h$) and $\mu_e$ ($\mu_h$) are the carrier density and mobility of electrons (holes), respectively. Fitting results yield $n_h = 3.30\times10^{21}$ cm$^{-3}$, $n_e = 1.79\times10^{21}$ cm$^{-3}$, $\mu_h = 78$ cm$^2\cdot$V$^{-1}\cdot$s$^{-1}$ and $\mu_e = 107$ cm$^2\cdot$V$^{-1}\cdot$s$^{-1}$. However, as the thickness decreased further, a linear Hall effect response emerged in the pristine NiTe$_2$ film with 10 nm (red curve in Fig. 2(f)), indicating that one band dominants the transport behavior in thinner film. The above results indicate the band structure evolution with varying the films' thickness. The thicker film in our study is similar to the bulk material and the Dirac cone is located near the Fermi level and two-carrier contribution dominates the Hall effect, which is consistent with bulk material study[27]. With decreasing the film thickness, a previous angle-resolved photoemission spectroscopy study reported that the MBE growth ultra-thin films have a band topology evolution and films above five-layer (~ 2.7 nm) are in the Type-II Dirac semimetal phase. Compared to the bulk material, in the five-layer thin film the Dirac cone is shifted above the Fermi level due to the reduced dimensionality[6], which is the possible reason for the observation of the hole carrier dominating Hall effect in our 10 nm NiTe$_2$ film.

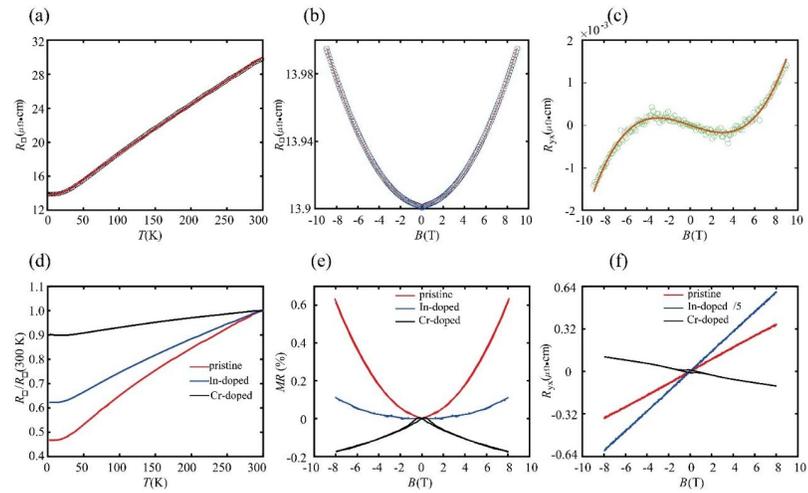

Fig. 2 (a-c) Electrical transport properties of 19.55 nm NiTe2 thin films. (a) The $RT$ curves at zero field, with the red line showing the BG equation fit. (b) $R_\square$ (B) curves at 2 K, with blue open circles for experimental data and the red line for the two-carrier model fit. (c) Hall resistivity curves at 2 K, with green open circles for experimental data and the red line for a two-carrier model fit. (d-f) Electrical transport properties of 10 nm NiTe$_2$ (red), Ni$_{0.99}$In$_{0.01}$Te$_2$(blue), and Ni$_{0.99}$Cr$_{0.01}$Te$_2$(black) thin films. (d) The $RT$ curve at zero field, with normalized resistance at 300 K. (e-f) $R_\square$(B) (e) and Hall resistivity (f) curves at 2 K.

Doping is widely used in semiconductors and topological insulators to modulate the electronic structure and magnetism of materials[29,30]. In and Cr have been used to substitute cations in topological insulators, achieving control over the band structure[31] and magnetism[21]. To study the modulation effect in thin $NiTe_2$ films via dilute elemental doping, Cr and In dopants are selected to substitute Ni sites. We prepared the $NiTe_2$ films by co-evaporating the doped elements during growth. The low doping concentration maintains the film's structure, as confirmed by the RHEED patterns as shown in the supplementary material Fig. S1. Figures 2(d)-(f) compare electrical transport properties of 10 nm films of pristine $NiTe_2$, $Ni_{0.99}In_{0.01}Te_2$, and $Ni_{0.99}Cr_{0.01}Te_2$, demonstrating significant modulation effects of elemental doping. The *RT* curves exhibit a notable resistance decreasing with decreasing temperature. In order to compare the *RT* relationship of these three samples, we plot the resistance as normalized at 300 K. The lowest normalized value drops from ~0.47 in the pure sample to 0.62 in the In-doped and 0.9 in the Cr-doped samples, indicating the existence of strong impurity scattering. The doped samples also show a significant decrease in magnetoresistance (*MR*), defined as $MR=(R(B)-R(0))/R(0)$, as seen in Fig. 2(e). The Cr-doped film exhibits negative *MR* with a butterfly-shaped curve, attributed to a ferromagnetic order. As shown in Fig. 2(f), Hall measurements reveal significant changes in carrier density: the In-doped sample retains hole dominance with decreasing density. In contrast, the Cr-doped sample changes towards the electron dominance. Consistent with the *MR* curve being indicative of the ferromagnetic emergence, the Hall effect reveals a square-shaped hysteresis loop. The above measurements indicate that Cr doping induces long-range ferromagnetic order in the films, whereas In doping does not. This can be understood as follows: Cr ions have unfilled *d*-electrons, resulting in a net magnetic moment, whereas In ions have fully filled *d*-electrons and cannot cause a net magnetic moment. We also performed the superconducting quantum interference devices (SQUID) characterization on the Cr-doped thin films and the magnetic order is confirmed. (See the detail in supplementary material Fig. S2). The above results highlight the substantial impact of dilute doping on the electrical transport and magnetic properties in $NiTe_2$ thin films.

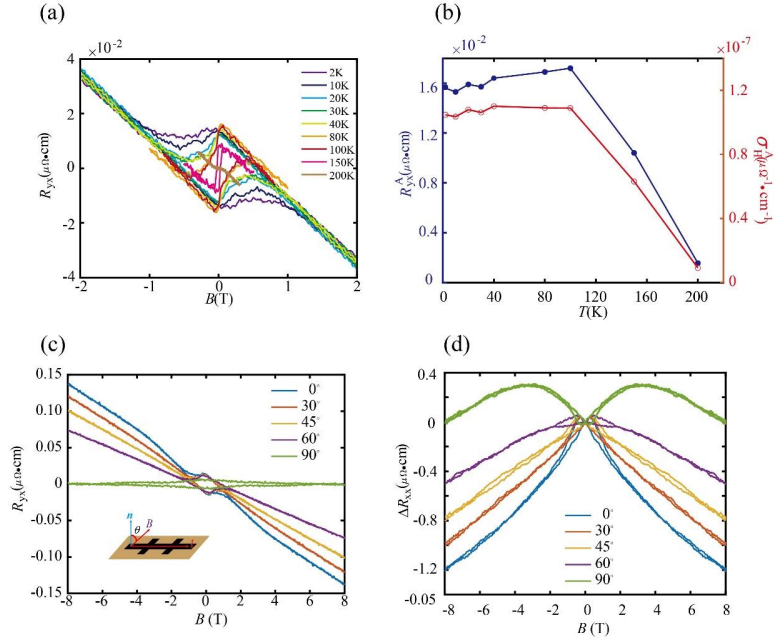

Fig. 3 Electrical transport properties of $Ni_{0.99}Cr_{0.01}Te_2$ thin film with a thickness of 10 nm. (a) Temperature dependent Hall resistivity curves with perpendicular magnetic field applied. (b) Temperature dependence of the anomalous Hall resistivity and conductivity. (c) Angle dependent Hall resistivity curves at 2 K, with the inset illustrating the transport measurement configuration with respect to the field orientation. (d) Angle dependent *MR* measurements at 2 K.

To clarify the origin of the anomalous Hall effect in Cr doped film, we measured the Hall effect across a wide range of temperature, as shown in Fig. 3(a). Assuming a simple model, the Hall effect has two contributions:

$$R_H = R_H^o + R_H^{AHE} \quad (5)$$

Where the first term on the right of the equation represents the ordinary Hall effect (OHE), determined by the carrier concentration, and the second term represents the anomalous Hall effect (AHE). All the Hall resistivity curves show a linear background with nearly identical slopes across temperatures, indicating that the carrier density is almost temperature-independent. The anomalous component's hysteresis loop become narrow with an increase in temperature and nearly disappeared at 200 K, which we regard as the Curie temperature. The SQUID measurement data is consistent with the transport measurement results (see Fig. S2 in the supplementary materials). This temperature is comparable to that of $Co_2Sn_3S_3$ compound, a typical magnetic Weyl semimetal. By extracting the data from Fig. 3(a), we find that the anomalous Hall conductivity remains nearly unchanged well below the Curie temperature, indicating an intrinsic band structure origin for the observed AHE. Additionally, a series of Hall resistivity and longitudinal resistivity were measured at various fixed tilt angles, $\theta$, between the external $B$ and normal direction to the substrate (illustrated in the inset of Fig. 3(c)). The results show that the anomalous Hall resistance changes negligibly small with varying the angle and is only suppressed near $\theta = 90°$, suggest that the film has a visible out-of-plane anisotropy magnetization, consistent with the SQUID magnetization results under both out-of-plane and in-plane magnetic fields. Regarding the longitudinal resistivity, $R_{xx}$ (where $\Delta R_{xx} = R_{xx}(B) - R_{xx}(0)$ in fig. 3(d)), we can see that almost all the films show a negative sign, which indicate a magnetic scattering suppression by magnetization. However, near $\theta = 90°$, the $\Delta R_{xx}$ turns to be positive and this seems to be related to the more gradual changes and disorder of magnetization. The angle-dependent $\Delta R_{xx}$ does not demonstrate the typical chiral anomaly characteristic of Weyl semimetals, which typically results in an extra conductivity contribution, leading to a negative magnetoresistance when the current and $B$ are non-perpendicular with each other. This absence could stem from the Fermi level remaining distant from the Weyl nodes, while contributions from additional bulk bands may mitigate the appearance of the chiral anomaly in the transport behavior. Further accurate energy band engineering via static gating or more precise doping may be required to address this issue.

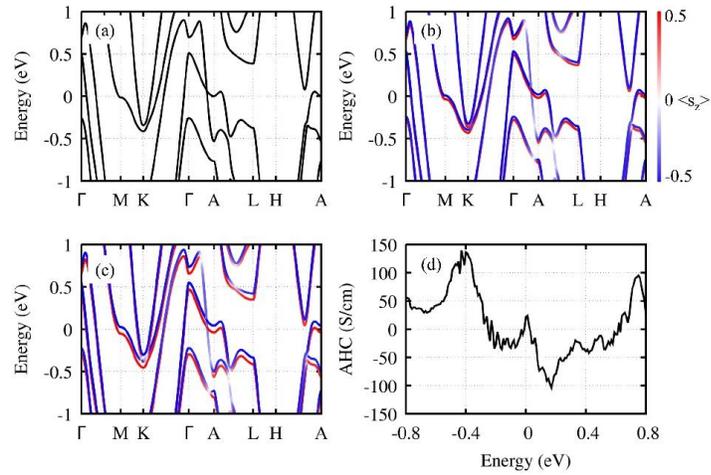

Fig. 4 (a) Electronic band structure of pristine NiTe$_2$ along high symmetry lines. (b-c) Electronic band structure of NiTe$_2$ with Zeeman field of 0.02 $\mu_B$ and 0.04 $\mu_B$, respectively. The red and blue color represent the spin direction. (d) Energy-dependent anomalous Hall conductivity with the consideration of spin splitting, corresponding to electronic band structure in (c).

To understand the mechanism of the AHE in the Cr- doped sample, we made a systematically first principles calculations for the evolution of band structures and corresponding intrinsic anomalous Hall conductivity[32]. Fig. 4 shows the energy dispersion of NiTe$_2$ along high symmetry lines. Owing to the $C_3$ rotation symmetry with respect to the z-axis, one pair of type-II Dirac points[33-35] locate on the Γ-A direction in the whole Brillouin zone near the Fermi level, as seen in Fig. 4(a). To see the magnetization effect on the band structure, we projected the Bloch wavefunction into maximally localized Wannier functions[36] and include the Zeeman field into the effective model Hamiltonians. From Figs. 4(b) and (c), one can see that each fourfold degenerated Dirac point splits into one pair of doubly degenerated Weyl points after the consideration of magnetization, and the spin splitting increases along with the magnitude of the Zeeman field. With consideration of magnetization along the z-direction, a non-zero finite anomalous Hall conductivity appears, as seen in Fig. 4(d). From energy-dependent anomalous Hall conductivity, one can easily see an obvious peak near the Fermi level. Interestingly, the position of this peak in energy space is almost equal to the location of the Weyl points. So, it implies that the anomalous Hall effect in magnetic doped $NiTe_2$ originated from the topological phase transition from Dirac semimetal to Weyl semimetal. Therefore, based on previous reports[6] and general topological band structure theory[37], we believe that type-II Dirac semimetal in $NiTe_2$ can transform to Weyl semimetal via magnetic doping.

In summary, as a type-II topological semimetal, $NiTe_2$ has rarely been researched in its low-dimensional forms. We prepared high-quality $NiTe_2$ thin films using MBE. In thicker films, $NiTe_2$ exhibited properties similar to its bulk material. However, in thinner films, electrical transport studies revealed significant changes in its band structure. Notably, co-deposition of dopant elements could enhance modulation effects. Specifically, Cr doping resulted in a visible out-of-plane anisotropic magnetization signal, and its anomalous Hall effect remained nearly constant across a wide temperature range, potentially due to the intrinsic topological nature of the band structure. Theoretical calculations further supported that introducing magnetic order into $NiTe_2$ could transform it into a magnetic Weyl semimetal. These results provide an important pathway and reference to the intriguing quantum states in low-dimensional $NiTe_2$ and its further developments in quantum devices.

**SUPPLEMENTARY MATERIAL**

See supplementary material for more details on RHEED patterns, magneto-electrical transport and SQUID measurements.

**ACKNOWLEDGMENTS**

This work is supported by the National Natural Science Foundation of China (Grant Nos. 12204043, 12174027, 12304208), Innovation Program for Quantum Science and Technology (Grant No. 2023ZD0300500).

**AUTHOR DECLARATIONS**

**Conflict of Interest**

The authors have no conflicts to disclose.

**Author Contributions**

Liguo Zhang and Dapeng Zhao contribute equally to this work.

**Reference**

[1] S. Manzeli, D. Ovchinnikov, D. Pasquier, O. V. Yazyev, and A. Kis, Nat. Rev. Mater. **2** (8), 17033 (2017).

[2] Z. Wei, B. Li, C. Xia, Y. Cui, J. He, J. B. Xia, and J. Li, Small Methods **2** (11) (2018).

[3] Y. Zhang, Y. Yao, M. G. Sendeku, L. Yin, X. Zhan, F. Wang, Z. Wang, and J. He, Adv. Mater. **31** (41), 1901694 (2019).

[4] M. Nurmamat, S. V. Eremeev, X. Wang, T. Yoshikawa, T. Kono, M. Kakoki, T. Muro, Q. Jiang, Z. Sun, M. Ye, and A. Kimura, Phys. Rev. B **104**, 155133 (2021).

[5] C. Xu, B. Li, W. Jiao, W. Zhou, B. Qian, R. Sankar, N. D. Zhigadlo, Y. Qi, D. Qian, F.-C. Chou, and X. Xu, Chem. Mater. **30** (14), 4823-4830 (2018).

[6] J. A. Hlevyack, L.-Y. Feng, M.-K. Lin, R. A. B. Villaos, R.-Y. Liu, P. Chen, Y. Li, S.-K. Mo, F.-C. Chuang, and T. C. Chiang, npj 2D Mater. Appl. **5**, 40 (2021).

[7] B. Ghosh, D. Mondal, C.-N. Kuo, C. S. Lue, J. Nayak, J. Fujii, I. Vobornik, A. Politano, and A. Agarwal, Phys. Rev. B **100**, 195134 (2019).

[8] M. Aras, Ç. Kılıç, and S. Ciraci, Phys. Rev. B **101**, 054429 (2020).

[9] P. P. Ferreira, A. L. R. Manesco, T. T. Dorini, L. E. Correa, G. Weber, A. J. S. Machado, and L. T. F. Eleno, Phys. Rev. B **103**, 125134 (2021).

[10] M. Qi, C. An, Y. Zhou, H. Wu, B. Zhang, C. Chen, Y. Yuan, S. Wang, Y. Zhou, X. Chen, R. Zhang, and Z. Yang, Phys. Rev. B **101**, 115124 (2020).

[11] F. Zheng, X.-B. Li, P. Tan, Y. Lin, L. Xiong, X. Chen, and J. Feng, Phys. Rev. B **101**, 100505(R) (2020).

[12] J. Zhang and G. Q. Huang, J. Phys. Condens. Matter. **32** (20), 205702 (2020).

[13] B. Pal, A. Chakraborty, P. K. Sivakumar, M. Davydova, A. K. Gopi, A. K. Pandeya, J. A. Krieger, Y. Zhang, M. Date, S. Ju, N. Yuan, N. B. M. Schroter, L. Fu, and S. S. P. Parkin, Nat. Phys. **18**, 1228-1233 (2022).

[14] T. Le, R. Zhang, C. Li, R. Jiang, H. Sheng, L. Tu, X. Cao, Z. Lyu, J. Shen, and G. Liu, Nat. Commun. **15**, 2785 (2024).

[15] B. S. de Lima, R. R. de Cassia, F. B. Santos, L. E. Correa, T. W. Grant, A. L. R. Manesco, G. W. Martins, L. T. F. Eleno, M. S. Torikachvili, and A. J. S. Machado, Solid State Commun. **283**, 27-31 (2018).

[16] I. Kar and S. Thirupathaiah, Mater. Today: Proc. **65**, 70-73 (2022).

[17] R. P. Dulal, B. R. Dahal, A. Forbes, N. Bhattarai, I. L. Pegg, and J. Philip, J. Vac. Sci. Technol. B **37**, 042903 (2019).

[18] J. Shi, Y. Huan, M. Xiao, M. Hong, X. Zhao, Y. Gao, F. Cui, P. Yang, S. J. Pennycook, J. Zhao, and Y. Zhang, ACS Nano **14** (7), 9011-9020 (2020).

[19] S. Pan, M. Hong, L. Zhu, W. Quan, Z. Zhang, Y. Huan, P. Yang, F. Cui, F. Zhou, J. Hu, F. Zheng, and Y. Zhang, ACS Nano **16** (7), 11444-11454 (2022).

[20] B. Seredyński, Z. Ogorzałek, W. Zajkowska, R. Bożek, M. Tokarczyk, J. Suffczyński, S. Kret, J. Sadowski, M. Gryglas-Borysiewicz, and W. Pacuski, Cryst. Growth Des. **21** (10), 5773-5779 (2021).

[21] C. Z. Chang, J. Zhang, X. Feng, J. Shen, Z. Zhang, M. Guo, K. Li, Y. Ou, P. Wei, L. L. Wang, Z. Q. Ji, Y. Feng, S. Ji, X. Chen, J. Jia, X. Dai, Z. Fang, S. C. Zhang, K. He, Y. Wang, L. Lu, X. C. Ma, and Q. K. Xue, Science **340**, 167-170 (2013).

[22] M. Mogi, M. Kawamura, R. Yoshimi, A. Tsukazaki, Y. Kozuka, N. Shirakawa, K. S. Takahashi, M. Kawasaki, and Y. Tokura, Nat. Mater. **16**, 516-521 (2017).

[23] E. Liu, Y. Sun, N. Kumar, L. Muchler, A. Sun, L. Jiao, S. Y. Yang, D. Liu, A. Liang, Q. Xu, J. Kroder, V. Suss, H. Borrmann, C. Shekhar, Z. Wang, C. Xi, W. Wang, W.

Schnelle, S. Wirth, Y. Chen, S. T. B. Goennenwein, and C. Felser, Nat. Phys. **14**, 1125-1131 (2018).

[24]D. Liu, A. Liang, E. Liu, Q. Xu, Y. Li, C. Chen, D. Pei, W. Shi, S. Mo, and P. Dudin, Science **365**, 1282-1285 (2019).

[25]T. Song, R. Bachelet, G. Saint-Girons, R. Solanas, I. Fina, and F. Sánchez, ACS Appl. Electron. Mater. **2** (10) 3221–3232 (2020)

[26]Y.-L. Ma, K. Zhu, and M. Li, Phys. Chem. Chem. Phys. **20** (43), 27539-27544 (2018).

[27]Q. Liu, F. Fei, B. Chen, X. Bo, B. Wei, S. Zhang, M. Zhang, F. Xie, M. Naveed, X. Wan, F. Song, and B. Wang, Phys. Rev. B **99**, 155119 (2019).

[28]C.-Z. Li, J.-G. Li, L.-X. Wang, L. Zhang, J.-M. Zhang, D. Yu, and Z.-M. Liao, ACS nano **10** (6), 6020-6028 (2016).

[29]G. Kresse and J. Hafner, Phys. Rev. B **48**, 13115 (1993).

[30]M. Ilegems, R. Dingle, and L. W. Rupp Jr., J. Appl. Phys. **46**, 3059–3065 (1975).

[31]R. Yu, W. Zhang, H. -J. Zhang, S. -C Zhang, X. Dai, and Z. Fang, Science, **329**, 61-64 (2010).

[32]L. Wu, M. Brahlek, R. Valdés Aguilar, A. V. Stier, C. M. Morris, Y. Lubashevsky, L. S. Bilbro, N. Bansal, S. Oh, and N. P. Armitage, Nat. Phys. **9**, 410 (2013).

[33]T. Zhang, Y. Jiang, Z. Song, H. Huang, Y. He, Z. Fang, H. Weng, and C. Fang, Nature **566**, 475-479 (2019).

[34]F. Tang, H. C. Po, A. Vishwanath, and X. Wan, Nature **566**, 486-489 (2019).

[35]M. Vergniory, L. Elcoro, C. Felser, N. Regnault, B. A. Bernevig, and Z. Wang, Nature **566**, 480-485 (2019).

[36]N. Marzari and D. Vanderbilt, Phys. Rev. B **56**, 12847 (1997).

[37]Z. Wang, Y. Sun, X.-Q. Chen, C. Franchini, G. Xu, H. Weng, X. Dai, and Z. Fang, Phys. Rev. B **85**, 195320 (2012).